\documentclass[10pt,letterpaper]{article}
\usepackage{opex3}

\usepackage{amsmath,amssymb}
\usepackage{color}
\usepackage{amssymb}
\usepackage{amsmath}
\usepackage{stackrel}
\usepackage{subfigure}
\usepackage{hyperref}

\definecolor{Gray}{rgb}{0.5,0.5,0.5}
\definecolor{darkblue}{rgb}{0,0,0.7}

\newcommand{\changed}[1]{{\textcolor{black}{#1}}}

\newcommand{\vect}[1]{{\bf{#1}}}
\newcommand{\vectgreek}[1]{{\boldsymbol{#1}}}
\newcommand{\mat}[1]{{\bf{#1}}}

\newcommand{\lightf}{l}

\newcommand{\lightfV}{{\vect{\lightf}}}

\newcommand{\proj}{\mat{P}}
\newcommand{\projM}{{\bf{\proj}}}
\newcommand{\image}{i}
\newcommand{\imageV}{{\vect{\image}}}
\newcommand{\auxvarV}{{\lightfV}}
\newcommand{\vectorize}[1]{\mathrm{vec}(\,{#1})}
\newcommand{\vectorizeinv}[1]{\mathrm{ivec}(\,{#1}\,)}
\newcommand{\layerA}{{\mat{F}}}
\newcommand{\layerB}{{\mat{G}}}

\newcommand{\lagrangemult}{\vect{\lambda}}

\newcommand{\argmin}[1]{\stackrel[\{ #1 \}]{}{\textrm{arg}\,\textrm{min}}}
\newcommand{\minimize}[1]{\stackrel[\{ #1 \}]{}{\textrm{minimize}}}

\newcommand{\shortcite}[1]{{\cite{#1}}}

\begin{document}

\title{A Compressive Multi-Mode Superresolution Display}

\author{Felix Heide$^1$, James Gregson$^1$, Gordon Wetzstein$^2$, Ramesh Raskar$^2$ {\upshape and} \\ Wolfgang Heidrich$^{1,3}$}
\address{1: University of British Columbia, 2: MIT Media Lab\\3: King Abdullah University of Science and Technology}

\email{fheide@cs.ubc.ca, gordonw@media.mit.edu} 



\begin{abstract}
Compressive displays are an emerging technology exploring the
co-design of new optical device configurations and compressive
computation. Previously, research has shown how to improve the dynamic
range of displays and facilitate high-quality light field or
glasses-free 3D image synthesis. In this paper, we introduce a new
multi-mode compressive display architecture that supports switching
between 3D and high dynamic range (HDR) modes as well as a new super-resolution mode. The
proposed hardware consists of readily-available components and is
driven by a novel splitting algorithm that computes the pixel states
from a target high-resolution image. In effect, the display pixels
present a compressed representation of the target image that is
perceived as a single, high resolution image.
\end{abstract}

\bibliographystyle{osajnl}
\bibliography{SuperResolution}


\section{Introduction and Related Work}
\label{sec:introduction}

Throughout the last few years, display technology has undergone a
major transformation. Whereas improvements of display characteristics,
such as resolution and contrast, have traditionally relied exclusively
on advances in optical and electrical fabrication, computation has now
become an integral part of the image formation. Through the co-design
of display optics and computational processing, computational and
compressive displays have the potential to overcome fundamental
limitations of purely optical designs. Characteristics that can be
improved by a co-design of display optics and computation include
dynamic range~\cite{Seetzen:2004:HDRDisplay} and depth of field of
projectors~\cite{Grosse:10}. A significant amount of research has
recently been conducted on compressive light field display for
glasses-free 3D image
presentation~\cite{lanman2010content,wetzstein2011layered,lanman2011polarization,wetzstein2012tensor}.

There has also been considerable work on computational superresolution
displays in recent years~\cite{Platt:SubpixelRendering}. Many recently
proposed approaches integrate different images optically by
superimposing multiple projections on the same
screen~\cite{DameraVenkata:SuperResolutionDisplay,jaynes2003super}. This
can also be achieved in single display/projector designs using
mirrors~\cite{Allen:wobulation} or by displaying different patterns on
a quickly moving device~\cite{berthouzoz2012resolution,
Didyk:ResolutionEnhancement}. These approaches, however, have the need
for multiple projection devices and mechanically moving parts. One
solution that overcomes these problems is {\em Optical Pixel Sharing}
(OPS)~\cite{Sajadi:2012}, which uses two LCD panels and a `jumbling'
lens array in projectors to overlay a high-resolution edge image on a
coarse image to adaptively increase resolution. This way only the most strongest edges are improved adaptively, while smooth areas are
unchanged. This approach brings the problem of the decision of when an
edge should be superresolved or not, causing noticeable artifacts due
to the global edge threshold used in their method. It is not
immediately obvious how such approaches could be adapted from
projectors to flatpanel displays.

\begin{figure*}[h!]
\centering
	\includegraphics[width=\textwidth]{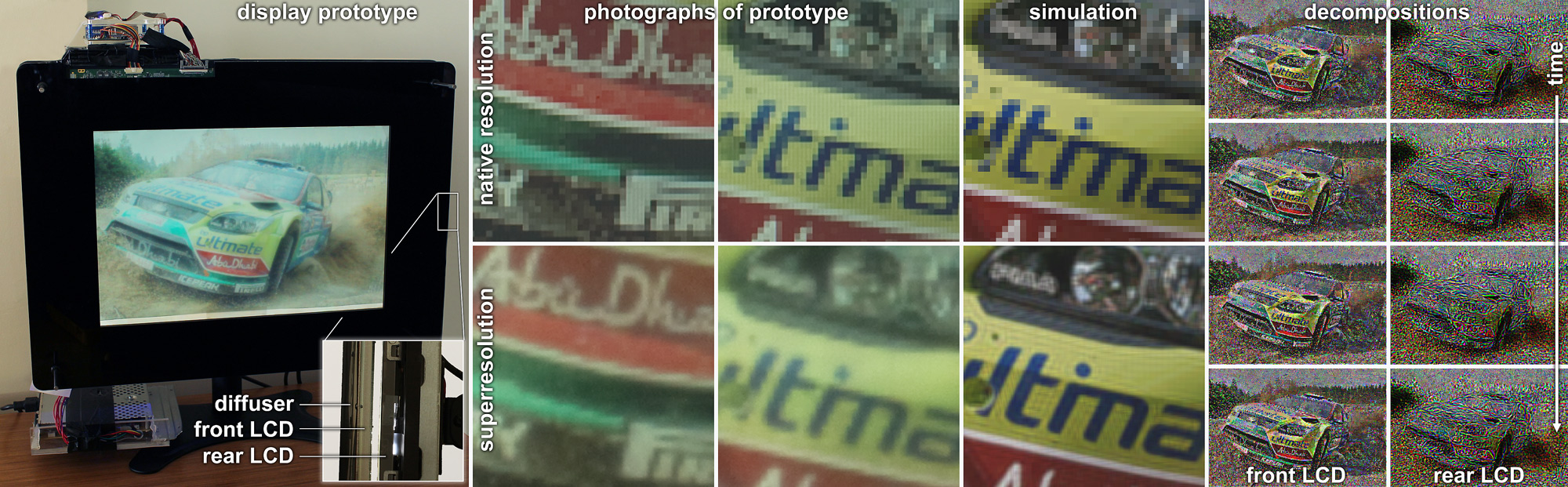}
\caption{Compressive superresolution display. The proposed display
  architecture comprises two stacked high-speed liquid crystal
  displays (LCDs) covered by a diffuser (left). A target
  high-resolution image is then decomposed into a set of patterns that
  are shown on the front and rear LCD in quick succession
  (right). Compared to the native resolution of each panel (center,
  top row), the proposed compressive display approach achieves
  significant improvements in resolution (center, bottom row) without
  any mechanically moving parts. Display content under CC license: \href{http://commons.wikimedia.org/wiki/File:Neste_Oil_Rally_2010_-_Jari-Matti_Latvala_in_shakedown.jpg}{LINK}}
\label{fig:teaser}
\end{figure*}

In this paper, we explore a new compressive display design that can be
switched between light field, high dynamic range, and superresolution
modes. The display design is inspired by previously proposed light
field displays~\cite{lanman2010content}, and is comprised of two
high-speed liquid crystal displays (LCDs) that are mounted in front of
each other with a slight offset (Fig.~\ref{fig:teaser}). To support a
super-resolution display mode, we introduce an additional diffuser
covering the LCD closest to the observer. The two stacked LCDs
synthesize an intermediate light field inside the device; the diffuser
then integrates the different views of that light field such that an
observer perceives a superresolved, two-dimensional image. Making the
diffuser electronically switchable allows for the display to be used
in 3D or high dynamic range mode.

\section{Superresolution Mode}
\label{sec:synthesis}

In this section, we derive models for superresolved image formation,
based on compressive light field displays. We also introduce inverse
methods to compute optimal pixel states for a target high-resolution
image.

\subsection{Image Formation}

\begin{figure}[t!]
	\centering
		\includegraphics[width=0.9\columnwidth]{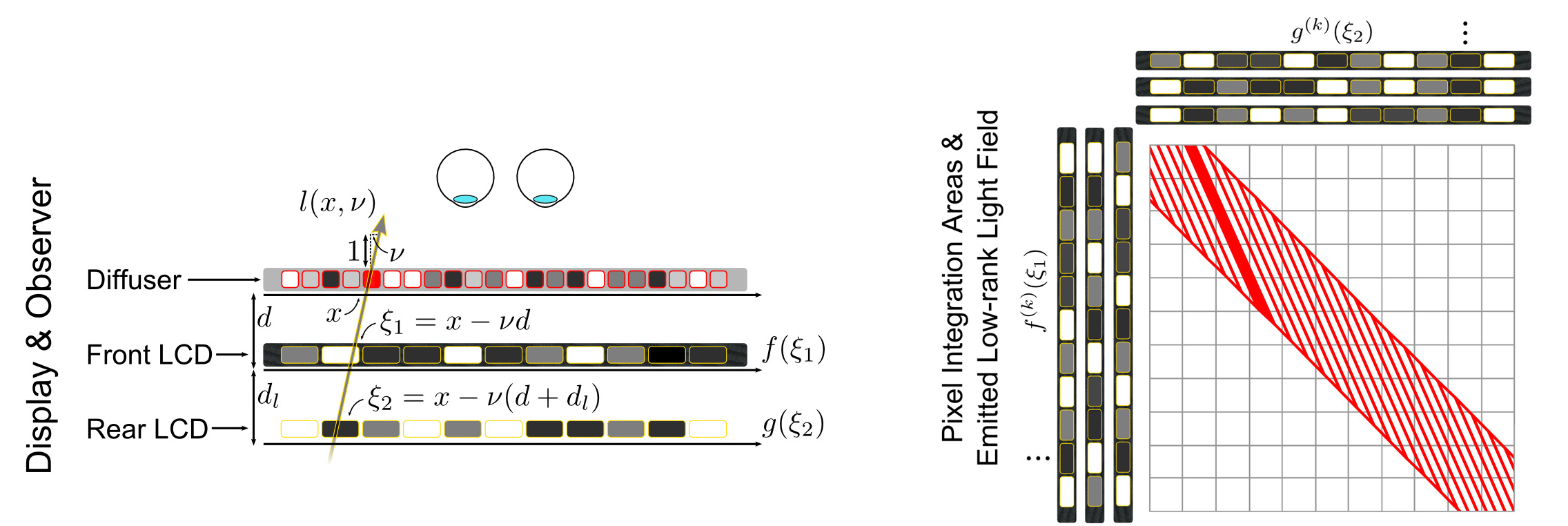}	
	\caption{Schematic of display components and parameters
          (Left). A diffuser is directly observed by the viewer and
          optically projects a 4D light field into a superresolved 2D
          image. The light field is emitted by two high-speed LCD
          panels. Optically, their combined effect is a multiplication
          allowing the light field to be represented by the outer
          product of their respective patterns $f(\xi_1)$ and
          $g(\xi_2)$ (Right). The pixels on the diffuser have a
          resolution exceeding that of either LCD panel, their
          integration areas are illustrated in red (Right).}
	\label{fig:coordinateSystem}
\end{figure}

The proposed optical display configuration comprises a light field display behind a diffuser. The image $i(\vect{x})$ observed on the diffuser is the integration of the incident light field $l( \vect{x}, \vectgreek{\nu})$ over the angular domain $\Omega_\vectgreek{\nu}$:
\begin{equation}
	i(\vect{x}) = \int_{\Omega_\vectgreek{\nu}} l( \vect{x}, \vectgreek{\nu}) \,d \vectgreek{\nu}.
\label{eq:integratedlightfield}
\end{equation}
Here, $\vect{x}$ is the 2D spatial coordinate on the diffuser and
$\vectgreek{\nu}$ denotes the angle. The light field is weighted with
angle-dependent integration weights of the diffuser. We employ a
relative two-plane parameterization of the light
field~\cite{durand2005frequency} (see Fig.~\ref{fig:coordinateSystem},
top). Conceptually, any light field display can be placed behind the
diffuser; we follow Lanman et al.~\shortcite{lanman2010content} and
use two stacked liquid crystal displays (LCDs). Driven at a speed
beyond the critical flicker frequency of the human visual system
(HVS), an observer perceives the temporal integral of the sets of
patterns shown on the display. The light field that is synthesized
inside the display and incident on the diffuser is 
\begin{equation}
	\tilde{l}( \vect{x}, \vectgreek{\nu}) = \frac{1}{K} \sum_{k=1}^K \, f^{(k)}( \vect{x} - d\cdot\vectgreek{\nu}) \cdot g^{(k)}( \vect{x} - ( d + d_l) \cdot\vectgreek{\nu}),
\label{eq:emittedlightfield}
\end{equation}
where $d$ is the distance between diffuser and front panel and $d_l$
is the distance between front and rear panel
(Fig.~\ref{fig:coordinateSystem}, top). The spatial coordinates on the
panels are denoted by $\vectgreek{\xi}$ whereas the functions $f(
\vectgreek{\xi_1})$ and $g( \vectgreek{\xi_2})$ give the transmittance
of front and rear panel at each position.

In this model, the panels run at a frame rate that is $K$ times faster
than the HVS. As observed by Lanman et
al.~\shortcite{lanman2010content}, the emitted light field of any pair
of LCD patterns corresponds to their outer product and is therefore
rank-1. The light field observed from the high-speed panels
$\tilde{l}( \vect{x}, \vectgreek{\nu})$ is rank-$K$ due to the retinal
integration of $K$ rank-1 light fields. Combining
Equations~\ref{eq:integratedlightfield} and~\ref{eq:emittedlightfield}
results in the following expression for the image observed on the diffuser:
\begin{eqnarray}
	\begin{aligned}
		\tilde{i}( \vect{x} ) 
		& = \int_{\Omega_\vectgreek{\nu}}  \frac{1}{K} \sum_{k=1}^K \, \left( f^{(k)}( \vect{x} - d \cdot\vectgreek{\nu} ) \cdot g^{(k)}( \vect{x} - ( d + d_l ) \cdot\vectgreek{\nu} ) \right)  d \vectgreek{\nu} \\
		& = \frac{1}{K} \! \sum_{k=1}^K \int \! \! \int \phi( \vect{x} \! - \! \vectgreek{\xi_{1}}, \vect{x} \! - \! \vectgreek{\xi_{2}}) \left( f^{(k)}( \vectgreek{\xi_1}) \cdot g^{(k)}( \vectgreek{\xi_2}) \right) d \vectgreek{\xi_{1,2}}. 
	\end{aligned}
	\label{eq:integratedemittedlightfield}
\end{eqnarray}
Equation~\ref{eq:integratedemittedlightfield} shows that each location on the diffuser integrates over some area on the front and rear LCDs. This integration is modeled as a convolution with a 4D kernel $\phi$. For an infinitely small point $\vect{x}$ on the diffuser, the kernel is
\begin{equation}
	\phi( \vectgreek{\xi_{1}}, \vectgreek{\xi_{2}}) = \mathrm{rect}( \vectgreek{\xi_1} / \vect{s_1})\, \mathrm{rect}( \vectgreek{\xi_2} / \vect{s_2} )\, \delta\hspace{-2pt}\left( \vectgreek{\xi_2} - \frac{\vect{s_2} }{\vect{s_1} }\vectgreek{\xi_1}\right),
	\label{eq:kernel}
\end{equation}
where $\vect{s_{1,2}}$ represent the spatial extent of the diffused
point on the front and rear panel, respectively, and $\mathrm{rect}(\cdot)$ is the rectangular function. These sizes depend on
the distance between the panels $d_l$, that between panel and diffuser
$d$, and the angular diffusion profile of the diffuser (see
Fig.~\ref{fig:coordinateSystem}).  In practice, the integration areas of
each superresolved pixel are calibrated for a particular display
configuration (see Sec.~\ref{sec:implementation}). Discretizing
Equation~\ref{eq:integratedemittedlightfield} results in
\begin{equation}
	\vect{i} = \mat{P} \, \vectorize{ \layerA  \layerB^T}.
\label{eq:integratedlightfield_discrete}
\end{equation}
Here, the $K$ time-varying patterns of front and rear LCD panels are encoded in the columns of matrices $\layerA  \in \mathbb{R}^{M \times K}$ and $\layerB \in \mathbb{R}^{M \times K}$, respectively (see Fig.~\ref{fig:coordinateSystem}), bottom). The resolution of the observed image $\vect{i} \in \mathbb{R}^{N}$ is larger than that of either panel, i.e. $N \geq M$. The convolution kernel is encoded in a discrete projection matrix $\mat{P} \in \mathbb{R}^{N \times M^2}$ and $\vectorize{\cdot}$ is a linear operator that reshapes a matrix into a vector by stacking up its rows.

Figure~\ref{fig:coordinateSystem} (right) illustrates the low-rank
light field matrix emitted by the two display layers along with the
integration areas of the superresolved pixels. Although each of these
is smaller than the regular grid cells of light rays spanned by the
display, the superresolved pixels are not aligned with the grid and
each pixel receives contributions from multiple different rays,
allowing for superresolution image synthesis.

\subsection{Superresolution Image Synthesis}

Given a target high-resolution image $\vect{i}$ and the image formation derived in the last subsection, we can formulate an objective function that minimizes the $\ell_2$-norm between the target and emitted images given physical constraints of the pixel states
\begin{equation}
	\begin{array}{cl}
		\minimize{\layerA , \layerB} & \left\| \vect{i} - \mat{P} \, \vectorize{ \layerA  \layerB^T } \right\|_2^2 \\
		\textrm{s.t.}  & 0 \leq \layerA, \layerB \leq 1
	\end{array}
	\label{eq:objective}
\end{equation}
This objective is difficult to deal with, as it involves a large matrix factorization embedded within a deconvolution problem. To make the problem manageable, we split the objective using the intermediate light field $\auxvarV$ produced by the display as a splitting variable
\begin{equation}
	\begin{array}{cl}
		\minimize{\layerA , \layerB} & \left\|  \layerA \layerB^T  - \vectorizeinv{\auxvarV} \right\|_F^2 \\
		\textrm{s.t.} & \projM \auxvarV  = \imageV, \; \, 0 \leq \auxvarV, \; \, 0 \leq \layerA, \layerB \leq 1 \\		
	\end{array}
	\label{eq:objective_split}
\end{equation}
Here, $\vectorizeinv{ \! \cdot \! }$ is a linear operator reshaping
the vector into a matrix, and the Frobenius norm $\| \! \cdot \!
\|_F^2$ measures the sum of squared differences of all matrix
elements. Although the objective function is non-convex, it is convex
with respect to each individual variable $\layerA, \layerB, \auxvarV$
with the other two fixed. \changed{The first constraint is affine in
  $\auxvarV$, an additional slack variable that splits the matrix
  factorization from the linear operator, while both are still
  coupled via the added consensus constraint.} We solve
Equation~\ref{eq:objective_split} using the alternating direction
method of multipliers (ADMM~\cite{boyd2011distributed}). 

After deriving the augmented Lagrangian for~\eqref{eq:objective_split}, the minimization of the augmented Lagrangian in each step leads to the following algorithm specific for our problem (see~\cite{boyd2011distributed} again for more details of the derivation of ADMM in general):
\begin{eqnarray}\label{eq:objective_split_update}
\begin{aligned}
  \auxvarV \leftarrow & \!\!
  \begin{array}[t]{cl}
		 \argmin{\auxvarV} & \mathcal{L}_\rho\left( \layerA , \layerB, \auxvarV, \lagrangemult \right) =
    \argmin{\auxvarV} \left\| \layerA \layerB^T \! - \vectorizeinv{
      \auxvarV } \right\|_F^2 + \rho \left\| \projM \auxvarV -
    \imageV + \vect{u} \right\|_2^2\\
    \textrm{s.t.} & 0 \leq \auxvarV
  \end{array}\!\!\!\!\!\!\!\!\!\!\!\\
  \left\{ \layerA, \layerB\right\} \leftarrow & \!\!
  \begin{array}[t]{cl}
		\argmin{\layerA, \layerB} & \mathcal{L}_\rho\left( \layerA , \layerB, \auxvarV, \lagrangemult \right) = 
    \argmin{\layerA, \layerB} \left\| \layerA \layerB^T -
    \vectorizeinv{ \auxvarV }\right\|_F^2 \\
    \textrm{s.t.} & 0 \leq \layerA, \layerB \leq 1
  \end{array}\\
  \vect{u} \leftarrow & \, \vect{u} + \big( \projM \auxvarV - \imageV
  \big)
\end{aligned}
\end{eqnarray}
\changed{where $\vect{u} = 1/\rho * \lagrangemult$ is a standard
  substitution that simplifies the notation~\cite{boyd2011distributed}.}

Using ADMM allows Equation~\ref{eq:objective_split} to be transformed into a
sequence of simpler subproblems. The first step of
Equation~\ref{eq:objective_split_update} is a deconvolution problem,
which we solve using SART iterations~\cite{yan2010convergence},
while the second step is a matrix factorization problem similar to the
one in the work by Lanman et al.~\shortcite{lanman2010content}. We
note that by splitting the objective in this manner, different light
field display technologies could be employed and would only require
the second term in the objective function to be replaced with the
appropriate image formation and inversion model.

\section{3D and High Dynamic Range Modes}
\label{sec:displaymodes}

As mentioned, we aim for a display architecture that can be operated
in both a superresolution mode as well as 3D and HDR modes. The
difference between these modes is the presence of the diffuser, which
should therefore be made electronically switchable for a multi-mode
display~\cite{gross2003blue,izadi2008going}.

\begin{figure}[h!]
\centering      
\includegraphics[height=4.4cm]{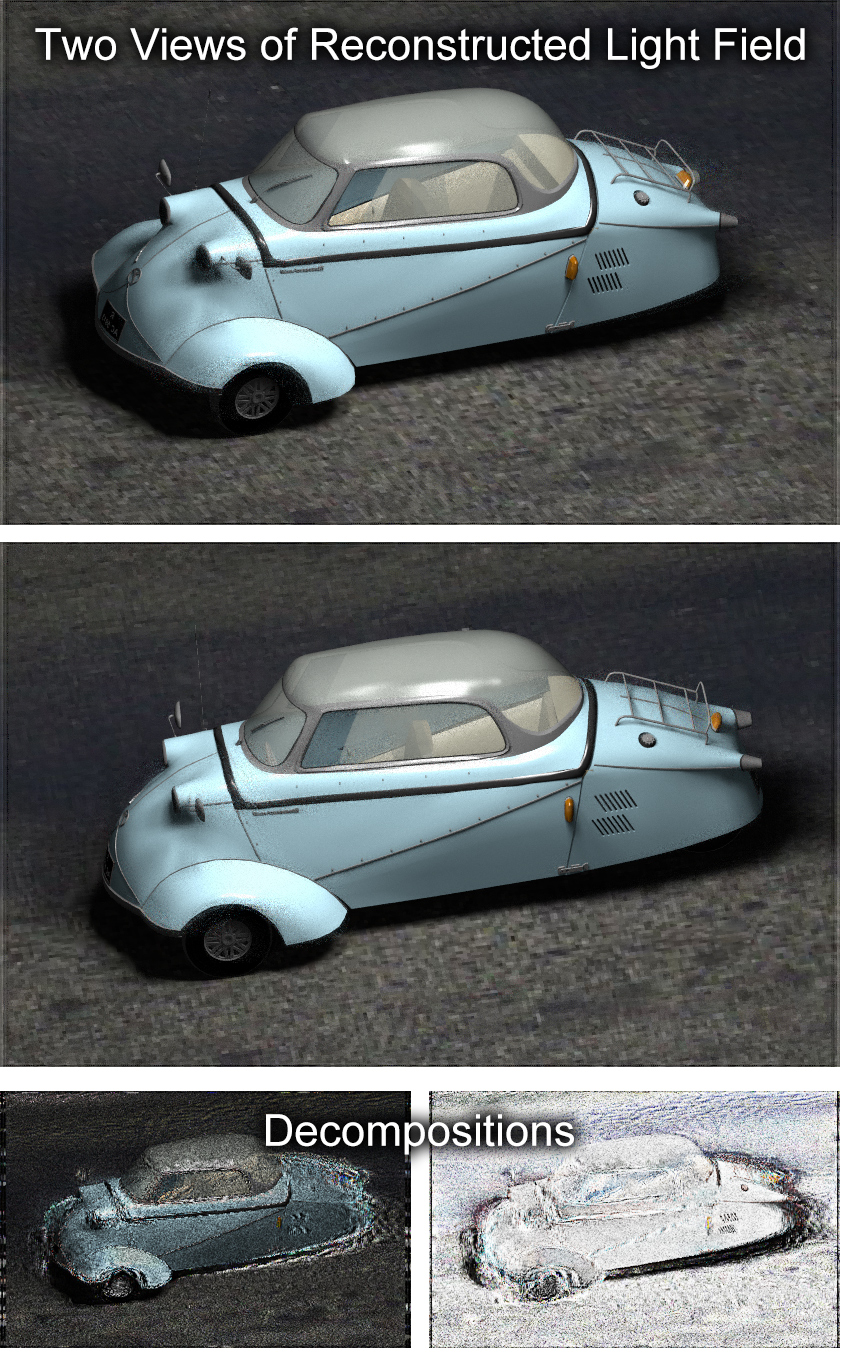} \hspace{0.5cm}
\includegraphics[height=4.4cm]{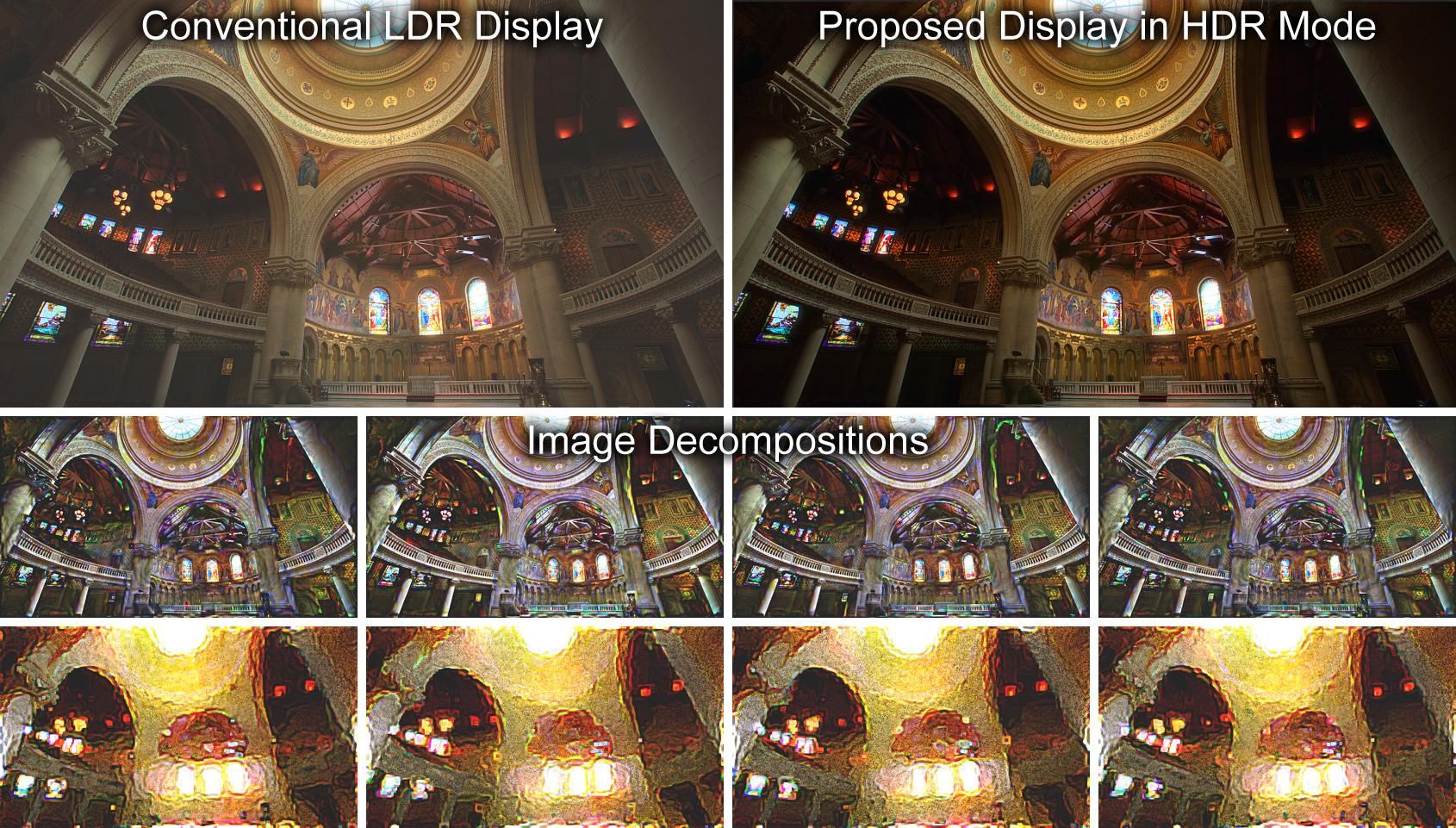}
\caption{\textbf{Left: Compressive light field display mode}. We can use the algorithm
by Lanman et al.~\shortcite{lanman2010content} to generate low-rank
glasses-free 3D content with the diffuser switched off. Simulated here is a rank-8 light
field that has $5 \times 3$ views. On the bottom we show a single time frame for both the
front and the rear liquid crystal panel. \textbf{Right: High dynamic range display mode.} The diffuser is
	switched off and a light field with no angular variation, but
	a higher contrast than that of the LCD panels is emitted. With
	a simulated black level of 15\%, a conventional display only achieves a low dynamic range (top left) whereas the proposed dual layer display (4-frame time-multiplexed here) significantly increases the dynamic range for 2D image display (top right). }
\label{fig:HR3D}
\end{figure}

With this diffuser switched off, the display hardware turns into a
duallayer light field display that is functionally equivalent to the
hardware configuration used by Lanman et
al.~\shortcite{lanman2010content}. We can therefore use the same
algorithms proposed in that work to produce glasses-free 3D content.
Figure~\ref{fig:HR3D} shows a simulation of the 3D mode with the same
parameters (LCD size, resolution, and spacing) as in the
superresolution mode.

As first pointed out by Wetzstein et
al.~\shortcite{wetzstein2011layered}, a multi-layer display can also
be used to represent a 2D image with an increased dynamic range
compared to the maximum available contrast on either the front or the
back panel. Whereas Wetzstein et al. rely on a tomographic image
formation model to generate HDR imagery, we directly apply a low-rank
approximation using the proposed mathematical framework. 

Figure~\ref{fig:HR3D} on the right illustrates the approach by
simulating a black level of 15\% of the peak intensity for both of the
two LCD panels. When presenting a conventional 2D image, its contrast
is severely reduced (top left). Using low-rank factorization, the
contrast of the displayed 2D image is significantly increased (top
right). The rank-4 decompositions for both LCDs are shown in the
bottom rows.

\section{Analysis and Evaluation}
\label{sec:analysis}

In the following we analyze the different parameters of the display
design in simulation.

\subsection{Analyzing Device Parameters}

\paragraph*{Diffusion Kernel and Spacing}

\begin{figure}[htb!]
	\centering
		\includegraphics[width=0.4\columnwidth]{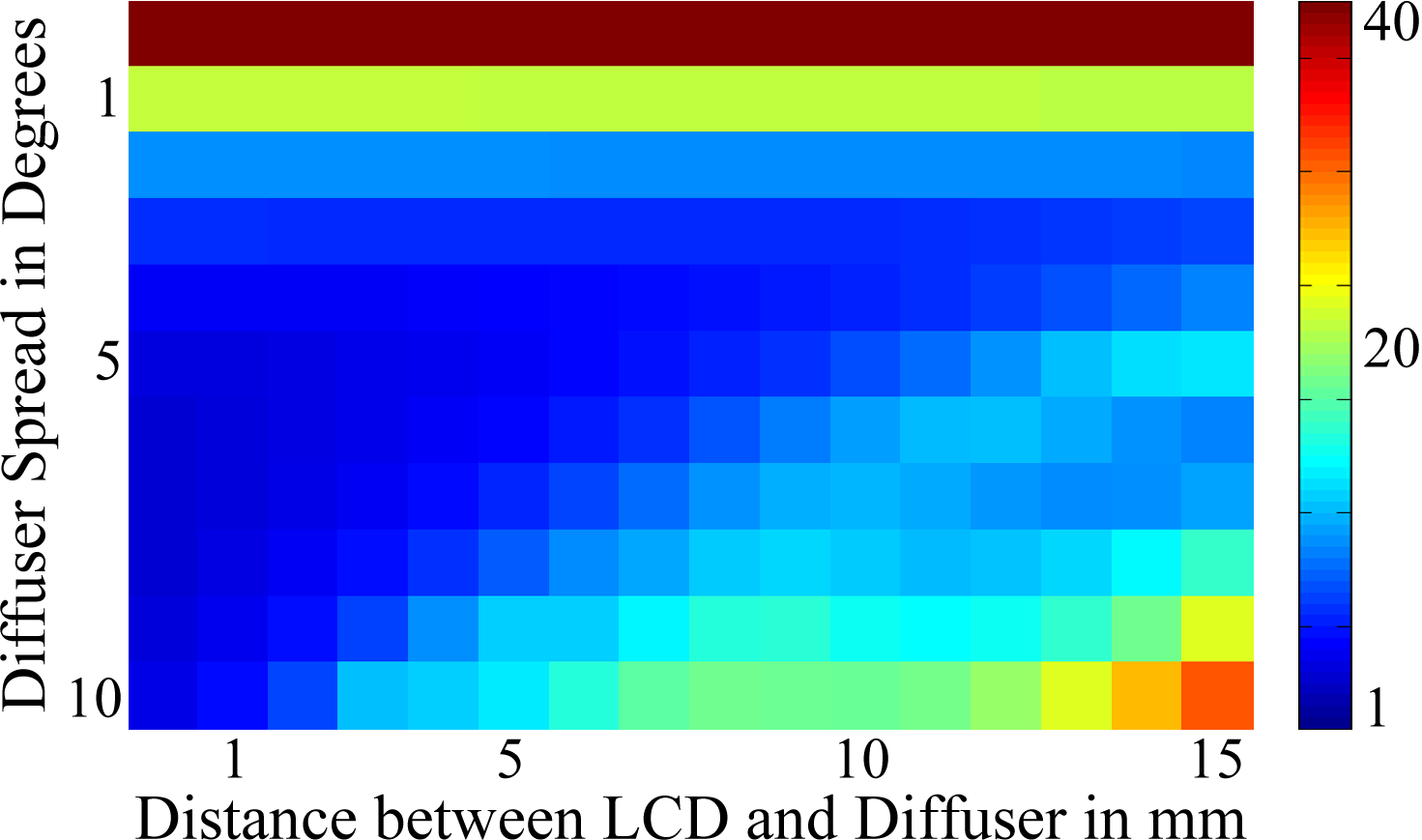}
		\caption{Conditioning analysis for a target 2$\times$ superresolution. We evaluate the condition number of the projection matrix for a varying distance between front LCD and diffuser as well as varying diffuser spread. A lower condition number corresponds to optical setups that are better suited for superresolution display. }
		\label{fig:AnalysisConditioning}
		\vspace{-0.3cm}
\end{figure}

In superresolution image acquisition techniques, the quality of an
optical setup is often analyzed using the condition number of the
projection matrix (e.g.,~\cite{Baker:Superresolution}). We follow this
approach: assuming that the proposed duallayer setup can synthesize
any light field, we plot the condition number of the projection matrix
$\mat{P}$ for a varying distance between front LCD panel
and diffuser as well as for a varying diffuser spread. Other display
parameters match those of the prototype described the Method section of our paper. Figure~\ref{fig:AnalysisConditioning}
shows the results for a target resolution increase of 2$\times$. 

A lower condition number corresponds to an optical configuration that is
better suited for inversion or, similarly, for superresolution
display. We observe that a very small distance with a diffusion spread
between 5 and 10 degrees results in the best expected image quality.

\vspace{-0.2cm}
\begin{figure*}[ht!]
\centering
	\includegraphics[width=\columnwidth]{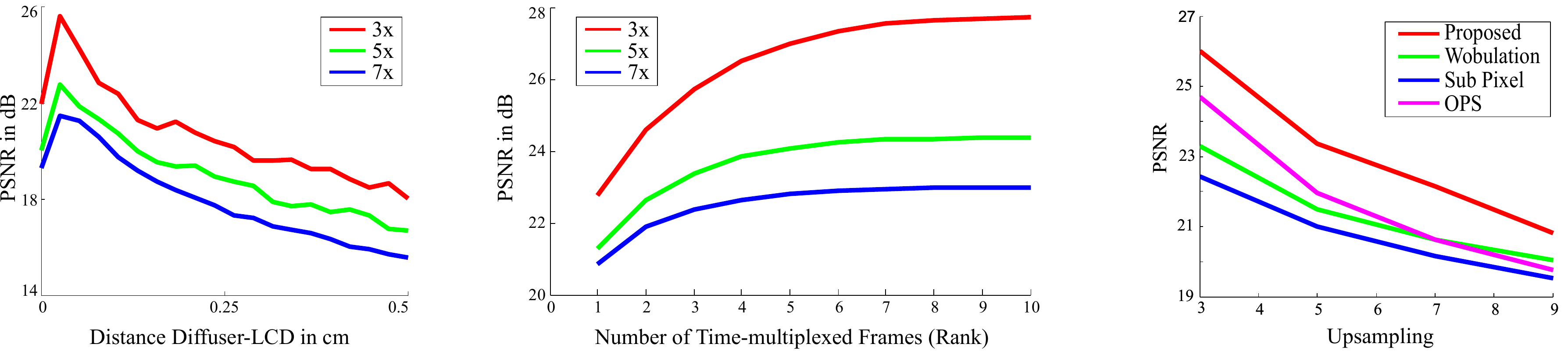}
\caption{Quantitative analysis of the proposed display
  architecture. Left: we simulate reconstructions of a test scene
  for a varying distance between front LCD and diffuser. The resulting
  image quality is best for a small distance. Center: reconstruction
  quality of a test scene on the prototype device is simulated for
  an increasing rank or number of subframes. While a
  higher rank allows for more degrees of freedom in the light field
  synthesis, only minor improvements in image quality are observed for
  ranks higher than six. Therefore, readily-available LCD panels with
  120 or 240 Hz are well-suited for computational superresolution
  display. Right: we compare the proposed method to \changed{Optical
    Pixel Sharing}, wobulation, and subpixel rendering for a varying
  superresolution factor.}
\label{fig:analysis}
\vspace{-0.3cm}
\end{figure*}

In Figure~\ref{fig:analysis} (left), we analyse the parameters
diffuser-spread, rank (time-multiplexing) and the upsampling factor
w.r.t. the superresolution performance using for simulations using a
set of natural images, which confirm the findings from the conditioning analysis. Using the prototype device
parameters, we simulate reconstructions of the scene from Fig.~\ref{fig:teaser} for
different superresolution factors. The number of subframes (rank) has
been fixed to 4 for this experiment to match the frame rate
capabilities of our prototype hardware. Both the PSNR analysis
and the scene-independent conditioning analysis are consistent: we observe that a small distance
between diffuser and front LCD is best suited for different
superresolution factors.

\paragraph*{Rank of the Light Field Factorization}

Faster LCD panels support more subframes, which in our system equates
to a higher rank approximation of the inverse problem. We therefore
expect the image quality to increase with the speed of the
panels. Assuming a critical flicker frequency of about 30 Hz, for
instance, allows two 120 Hz LCD panels to emit a rank-4 light field
because a human observer perceptually averages over four
time-multiplexed frames. We analyze the quality of superresolution
image generation for a varying light field rank in
Figure~\ref{fig:analysis} (center). In this experiment, the diffuser
spacing is fixed at 0.03 cm. For this example, we simulate
reconstructions of the scene from Fig.~\ref{fig:teaser} on the
prototype device. As seen in the plots, reconstruction quality using
our method asymptotically reaches a maximum for a given image as a
function of the light field rank. Only minor improvements are observed
for ranks higher than six. 

\vspace{-0.3cm}

\begin{figure}[h!]
\centering
\includegraphics[width=0.65\columnwidth]{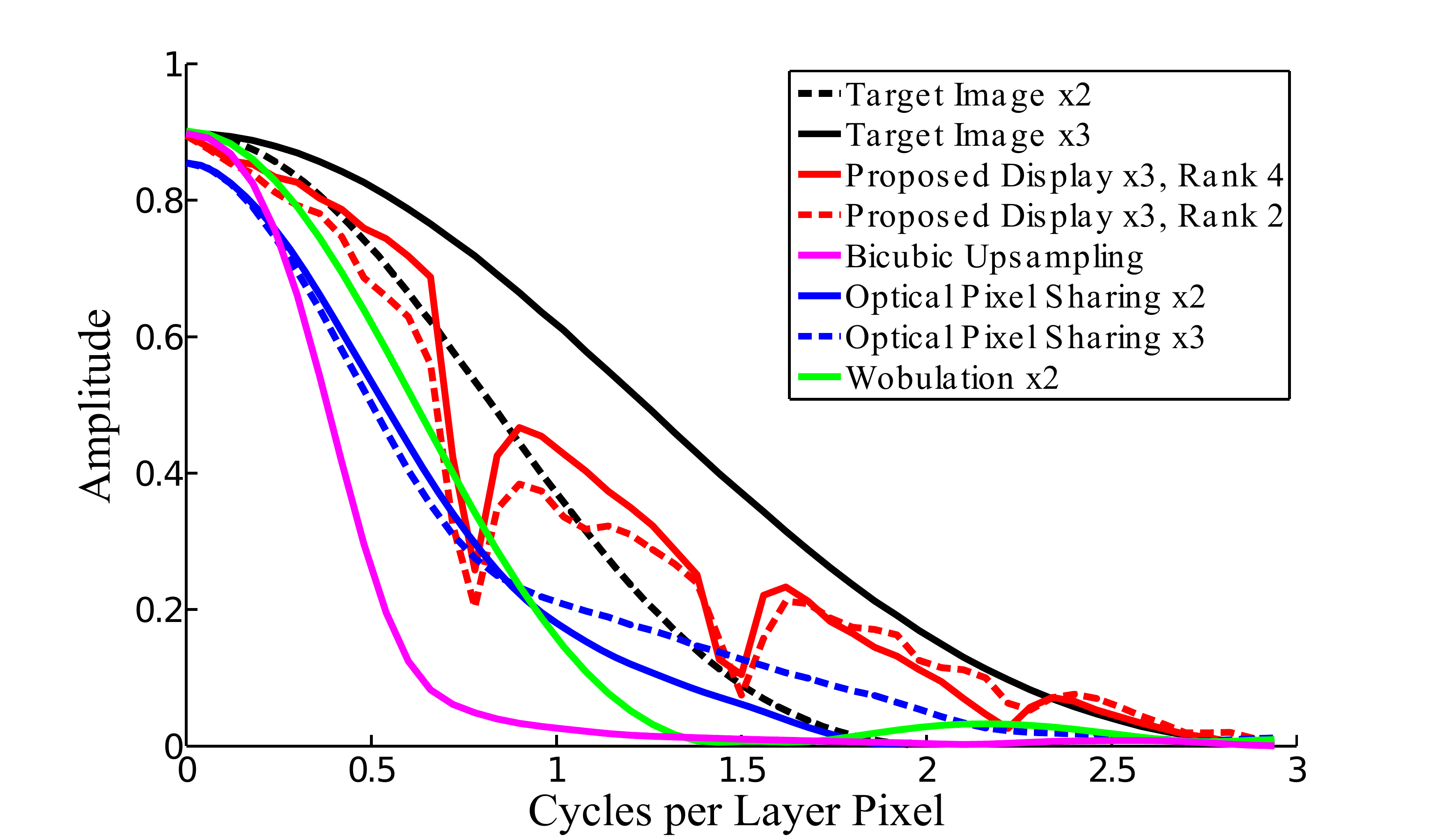}
\vspace{-0.1cm}
\caption{
  \changed{Quantitative resolution analysis for several different
    super-resolution displays based on simulated images for each
    system. We show the modulation transfer function (MTF) for each
    system, as measured with Burns' slanted edge
    method~\protect\cite{burns2000slanted}. The curves are normalized
    such that the Nyquist limit for the native hardware resolution is
    at 1. The black curves show slanted edge results for $2\times$ and
    $3\times$ superresolution target images. Our approach (red
    curves) generally preserves the most high frequencies, compared to
    other methods such as Optical Pixel Sharing and wobulation.}}
\label{fig:OPSComparison}
\vspace{-0.4cm}
\end{figure}

\paragraph*{Resolution Analysis}

Next, we perform a quantitative analysis is of the resolution
achievable with our approach. Due to the compressive nature of our
display design, the target image resolution used in the optimization
procedure is a goal that is
not, in general, achieved in the actual result. In order to obtain a
quantitative estimate of the actual resolution increase, we measure
the Modulation Transfer Function (MTF) of the display using the
slanted edge procedure Burns~\shortcite{burns2000slanted}.

The MTF curves for our display and a number of comparison methods are
shown in Figure~\ref{fig:OPSComparison}. All MTF curves are normalized
such that the Nyquist limit for the pixel size of one of the LCD
panels is at 1. The black curves show MTFs for $2\times2$ (dashed)
and $3\times3$ (solid) superresolution target images. For comparison we
also include a cubic upsampling filter (magenta), which performs a
sharpening operation and therefore contains slightly higher
frequencies than the Nyquist limit of the original pixel grid. Our
approach for a target $3\times3$ superresolution is shown in red for a
rank 4 display (solid) and for a rank 2 display (dashed). We note that both variants a resolution increase between a
factor of $2$ and $3$, with a diminishing resolution gain for higher
ranks.

\paragraph*{Comparison to Other Superresolution Displays}

Fig.~\ref{fig:OPSComparison} also provides a comparison with other
proposed superresolution display technologies.
The blue curves refer to Optical Pixel Sharing (OPS) with $2\times2$ (dashed) and $3\times3$ superresolution (solid). We note that in both cases, the OPS results are dominated by our approach, indicating that our design is overall better at reproducing higher frequencies
than OPS for the same resolution LCD panels. Another way to
compare the two methods is by considering parameter selections for
both methods that exhibit the same compressibility requirements,
i.e. the same ratio between degrees of freedom in the display hardware
vs. in the target image. The two dashed curves provide that
comparison, as both methods use the same number of subframes (2) for
the same superresolution target ($3\times3$).

Figure~\ref{fig:OPSComparison} shows that our method also outperforms
the wobulation approach (shown in green). With an equal number or half
the degrees of freedom (solid and dashed red lines, respectively)
available to wobulation, our method is able to successfully preserve
much of the high-frequency content up to the target 3 cycles per LCD
pixel.

Figure~\ref{fig:analysis} (right) shows a different way to compare
with wobulation~\cite{Allen:wobulation},~\cite{Platt:SubpixelRendering} and OPS.

The target image for this experiment was an image showing a chirp that
contains a range of frequencies. We see graceful degradation of image
quality as the superresolution factor is increased. Simulated results
from our method at a superresolution factor of three exceed the
reconstruction quality of the other two methods by a factor of two.

\section{Prototype}
\subsection{Methods}
\label{sec:implementation}

\paragraph{Display}
A superresolution display prototype was constructed from two Viewsonic
VX2268wm 120 Hz LCD panels. All diffusing and polarizing films were
removed from the front panel. The front-most (diffusing) polarizer was
replaced by a clear linear polarizer. Mounted on a rail system, the
panels were adjusted to have a spacing of 19 mm in between. The rear
panel had an unmodified backlight that illuminated the duallayer
device. Fixated to a frame that was adjustable on the rail system, the
diffuser was mounted at a distance of 6 mm to the front LCD.

The prototype was controlled by a 3.4 GHz Intel Core i7 workstation
with 4 GB of RAM. A four-head NVIDIA Quadro NVS 450 graphics card
synchronizes the two displays and an additional external monitor. With
the diffuser in place, the display functioned in superresolution mode
using content generated by the algorithm discussed in
Section~\ref{sec:synthesis}. With the diffuser removed, the display
functions in glasses-free 3D or HDR modes. While
electronically-switchable diffusers are commercially available, we did
not use one in our experiments.

\paragraph{Calibration}
For the prototype display, we calibrated display gamma curves,
geometric alignment of the LCD panels, and diffuser point spread
function (PSF).

\vspace{-0.4cm}
\begin{figure}[h!]
\centering
\includegraphics[width=0.35\columnwidth]{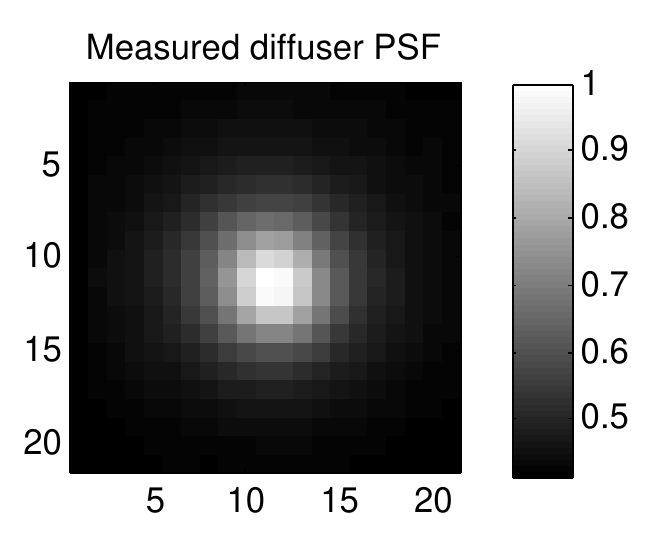}\hspace{0.2cm}
\includegraphics[width=0.35\columnwidth]{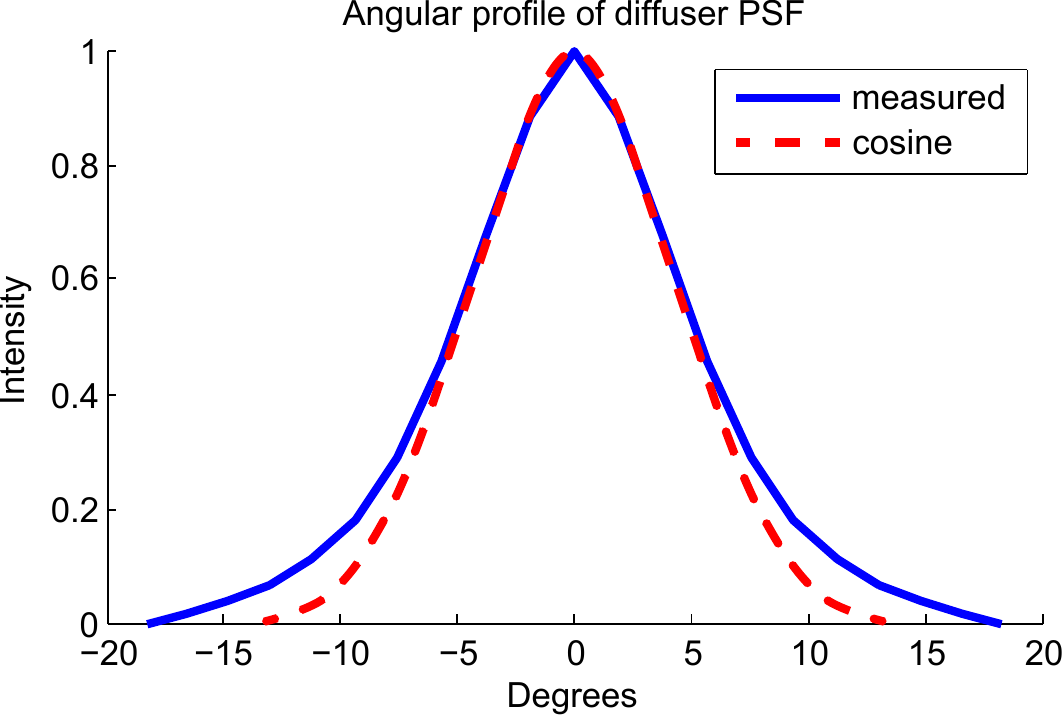}
\vspace{-0.2cm}
\caption{Measured diffusion point spread function of the prototype (left). The angular profile of the measured PSF is well-modeled by a rotationally-symmetric cosine function (right).}
\label{fig:psfcalib}
\end{figure}
\vspace{-0.1cm}

Gamma curves are calibrated using standard techniques: uniform images
with varying intensities are shown on the display and captured with a
linearized camera in RAW format. The acquired curves are inverted in
real-time when displaying decomposed patterns. The display black level
is incorporated as a constraint into the nonnegative matrix
factorization routine.

A second calibration step requires geometric registration of front and
rear LCDs. For this purpose, we aligned the panels mechanically as well
as possible and fine-tuned possible misalignments in software. With the
diffuser removed, we showed crossbars on both screens that were aligned
for the perspective of a calibration camera.

Finally, we measured the point spread function (PSF) of the diffuser by
displaying white pixels on a uniform grid on the front panel, with the
rear panel fully illuminated. The PSFs were then extracted from
linearized RAW photographs of the prototype by extracting areas around
the according grid positions. The PSFs measured on the prototype were
approximately uniform over the display surface, hence we averaged all
PSFs and used a spatially-invariant PSF in the computational
routines. Figure~\ref{fig:psfcalib} shows the calibrated PSF captured
from the device; this PSF is well modeled as a rotationally-symmetric
angular cosine function with a field of view of 15 degrees (shown on
the right).

\subsection{Results}
\label{sec:Results}

\paragraph{Simulated Results}

\begin{figure}[h!]
\centering
\includegraphics[width=0.82\columnwidth]{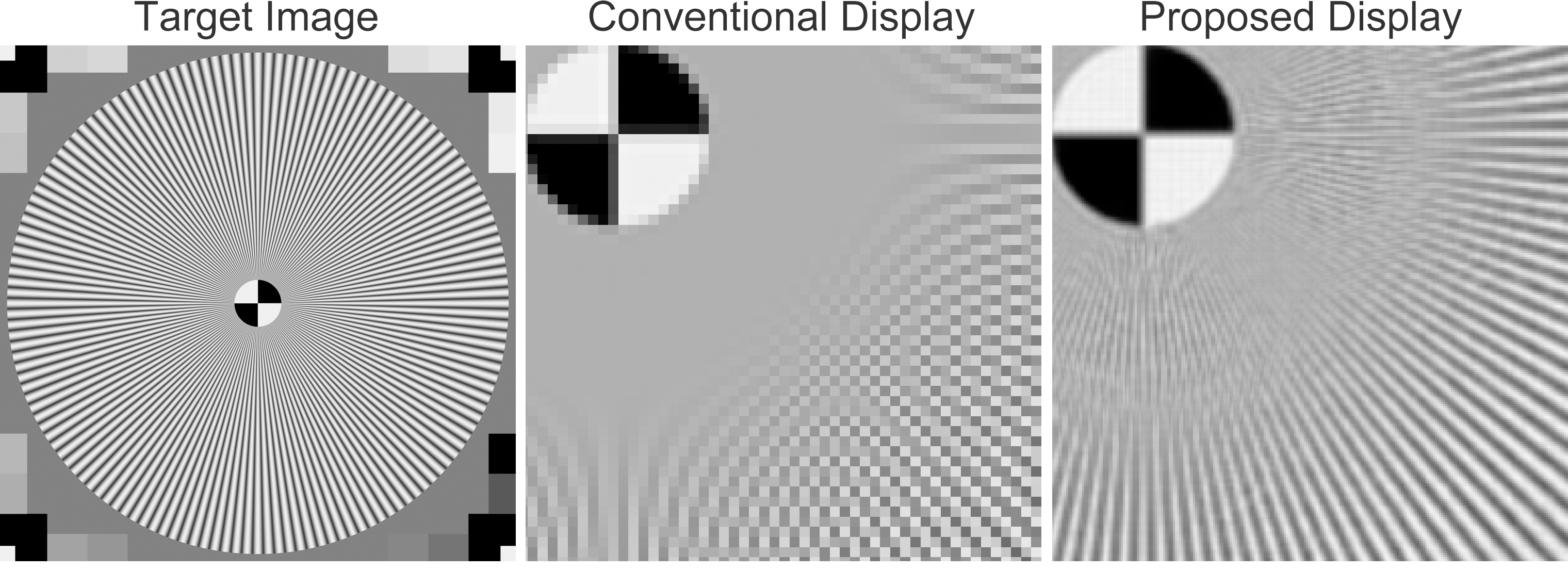}\vspace{0.5cm}
\includegraphics[width=0.82\textwidth]{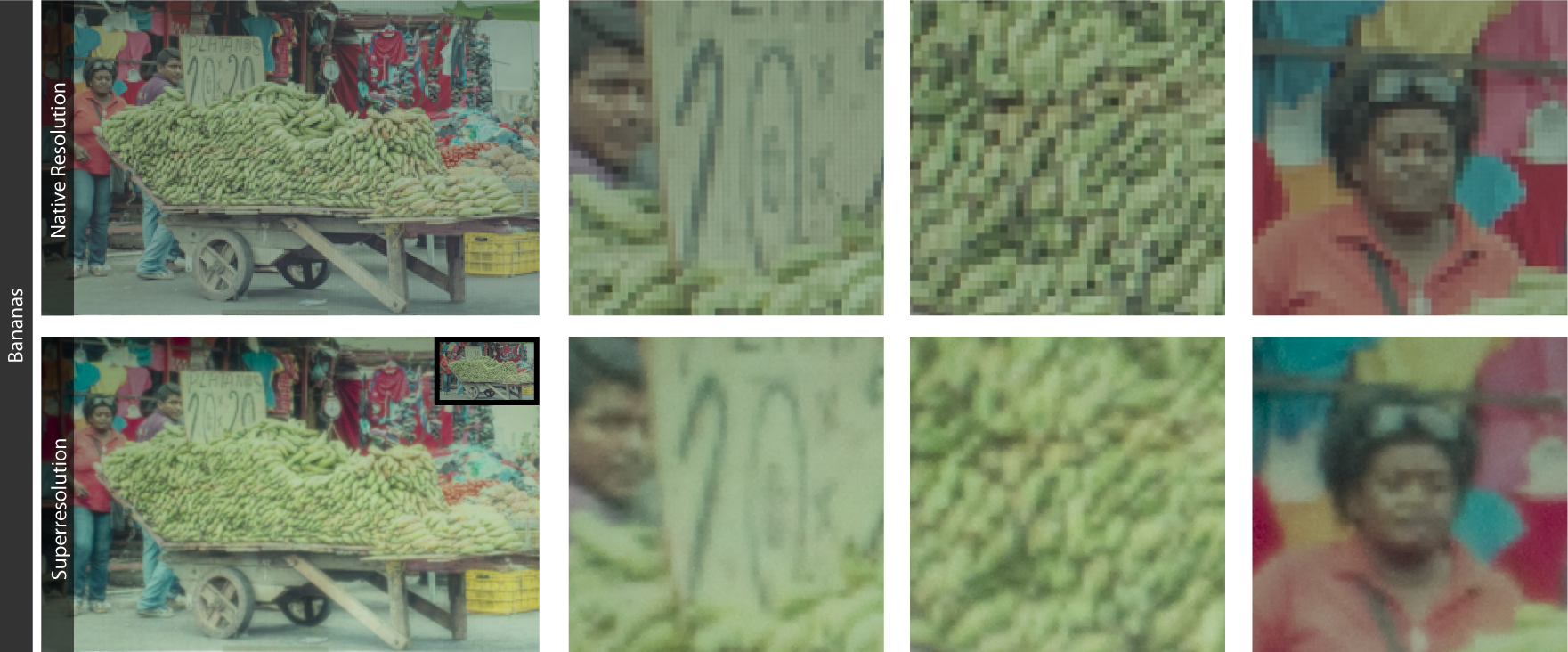}\vspace{0.2cm}
\includegraphics[width=0.82\textwidth]{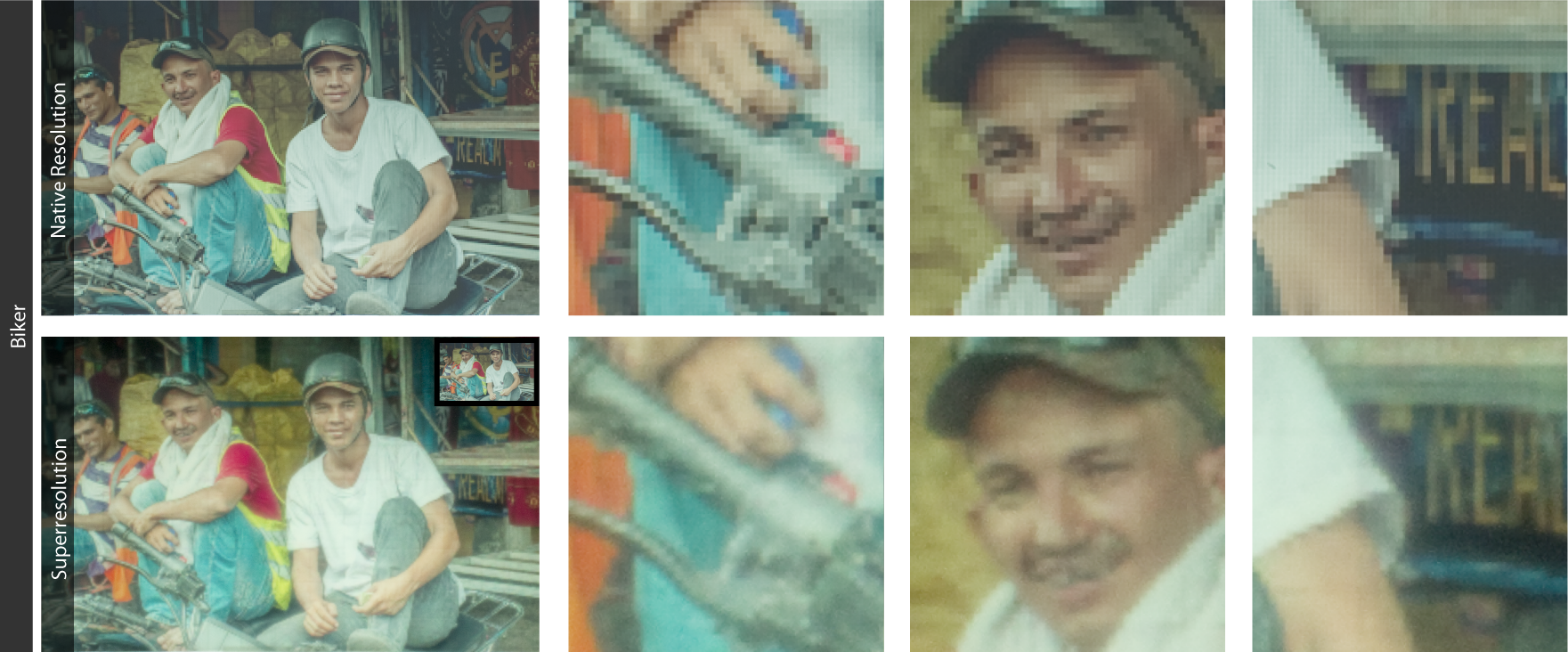}
\caption{\textbf{Top row:} Showing a high resolution target image (left) on a lower-resolution display results in a loss of image features (center). The proposed method is capable of preserving such features by displaying superresolved image content (right).
\textbf{Bottom and center row:} Photographs of prototype computational superresolution display. Each of the example scenes is superresolved at 5$\times$. The top row of each example shows photographs captured at the native display resolution while the bottom shows our method. Isolated and thin high-frequency details such as text on signs, butterfly antennae, and sharp off-axis edges are significantly enhanced by the proposed method. Faces are also dramatically improved as are textured regions such as the bananas, plant, and clothing. Display content under CC license:\href{http://commons.wikimedia.org/wiki/File:Since_selling_bananas_on_the_street.jpg}{LINK1}  \href{http://commons.wikimedia.org/wiki/File:Mototaxis.jpg}{LINK2}}
\label{fig:barb}
\end{figure}

Figure~\ref{fig:barb} (top) shows a simulated result for the proposed
superresolution display mode for a target upsampling factor of
5$\times$ and rank 4.

\paragraph{Results of Prototype Display}

The two bottom rows of Figure~\ref{fig:barb} show a result captured
using our prototype hardware. Images were captured by photographing
the display using an Canon T3 SLR camera. The bottom row shows a
result obtained at the nominal image resolution while the bottom row
shows superresolution results with a nominal upsampling of 5$\times$.

The example images have several faces and off-axis edges as well as sparse text
and textured regions.  All are improved on the prototype, including
increased detail on the brake handle, fine-scaled features around the
eyes and mouth of the face and dramatic improvements to the
pixel-scaled text.

\section{Discussion}
\label{sec:discussion}

In summary, we present a computational display approach to
superresolution image synthesis. Through the co-design of display
optics and computational processing, the proposed architecture is the
first to facilitate superresolution with a single device and without
the need for mechanically moving parts in a form factor suitable for
televisions and computer monitors. Moreover, it is the first design
that can be reconfigured on the fly between three radically different
display modes: superresolution, glasses-free 3D, and 2D high dynamic
range.

The off-the-shelf hardware components used in our prototype exhibit
limited contrast, color cross-talk, and interreflections between LCD
panels that result in slight deviations of observed results from
simulations; these deviations are perceived as ringing and loss of
contrast. While the employed 120 Hz LCD panels allow for rank-3 to
rank-4 decompositions, depending on the viewing conditions, higher
speed panels are desirable. Our simulations indicate that a refresh
rate of 240 Hz would be ideal for the application of superresolution
image display. Currently, we process each color channel separately and
do not take the panel-specific subpixel structures into
account. Diffraction has not been an issue in our experiments, but
significantly smaller pixel sizes would require such effects to be
included in the image formation. Finally, we envision the proposed
display to operate with an electronically-switchable diffuser but
leave this to future engineering efforts. 

\end{document}